\pdfoutput=1
\documentclass[doublecol]{epl2} 
\usepackage{color}
\usepackage{graphicx,amsmath}
\newcommand{\nn}{\nonumber}
\newcommand{\fig}[2]{\includegraphics[width=#1]{./figures/#2}}
\newcommand{\Fig}[1]{\includegraphics[width=\columnwidth]{./figures/#1}}

\newcommand{\sign}{{\mathrm{sign}}}

\newlength{\figsize}
\newcommand{\cm}[1]{}

\newcommand{\rme}{{\mathrm{e}}}
\newcommand{\rmd}{{\mathrm{d}}}

\begin{document}

%\bibliographystyle{KAY}
%\bibliographystyle{eplbib}
%%%%%%%%%%%%%%%%%%%%%%%%%%%%%%%%%%%%%%%%%%%%%%%%%%%%%%%%%%%%%%%%%%%%%%%%%%%%
\title{Fluctuation force exerted by a planar self-avoiding polymer}
\author{Pierre Le Doussal and Kay J\"org Wiese} 
\institute{CNRS-Laboratoire de Physique
Th{\'e}orique de l'Ecole Normale Sup{\'e}rieure, 24 rue Lhomond, 75231
Paris Cedex, France.}
\abstract{
Using results from Schramm L\"owner evolution (SLE),
we give the expression of the fluctuation-induced force
exerted by a polymer on a small impenetrable disk, in various 2-dimensional domain geometries.
We generalize to two polymers and examine whether the fluctuation force can trap the object into a stable equilibrium. We compute the force exerted on objects at the domain boundary, and the force
mediated by the polymer between such objects. The results can straightforwardly be extended to any SLE interface, including Ising, percolation, and loop-erased random walks.  Some are
relevant for extremal value statistics. 
}
\maketitle

What is the force exerted by a polymer on a small object, such as a mesoscopic disk or a molecule? Simply because the object cannot be penetrated by the polymer it constrains its thermal fluctuations and   feels an entropic force. This question is relevant in view of the recent  surge of interest in fluctuation-induced forces, such as Casimir forces, triggered by beautiful experiments in critical systems \cite{criticalcasimirexp}. Apart from Gaussian fluctuations, calculation of Casimir forces is difficult%\cite{casimirreview}
, and it is useful to obtain exact results for non-trivial theories \cite{criticalcasimirtheo}. 
\begin{figure}
\begin{center}
\leftline{\fbox{$w$}\hspace{4cm}{\fbox{$z$}}}\vspace*{-0.3cm}
\Fig{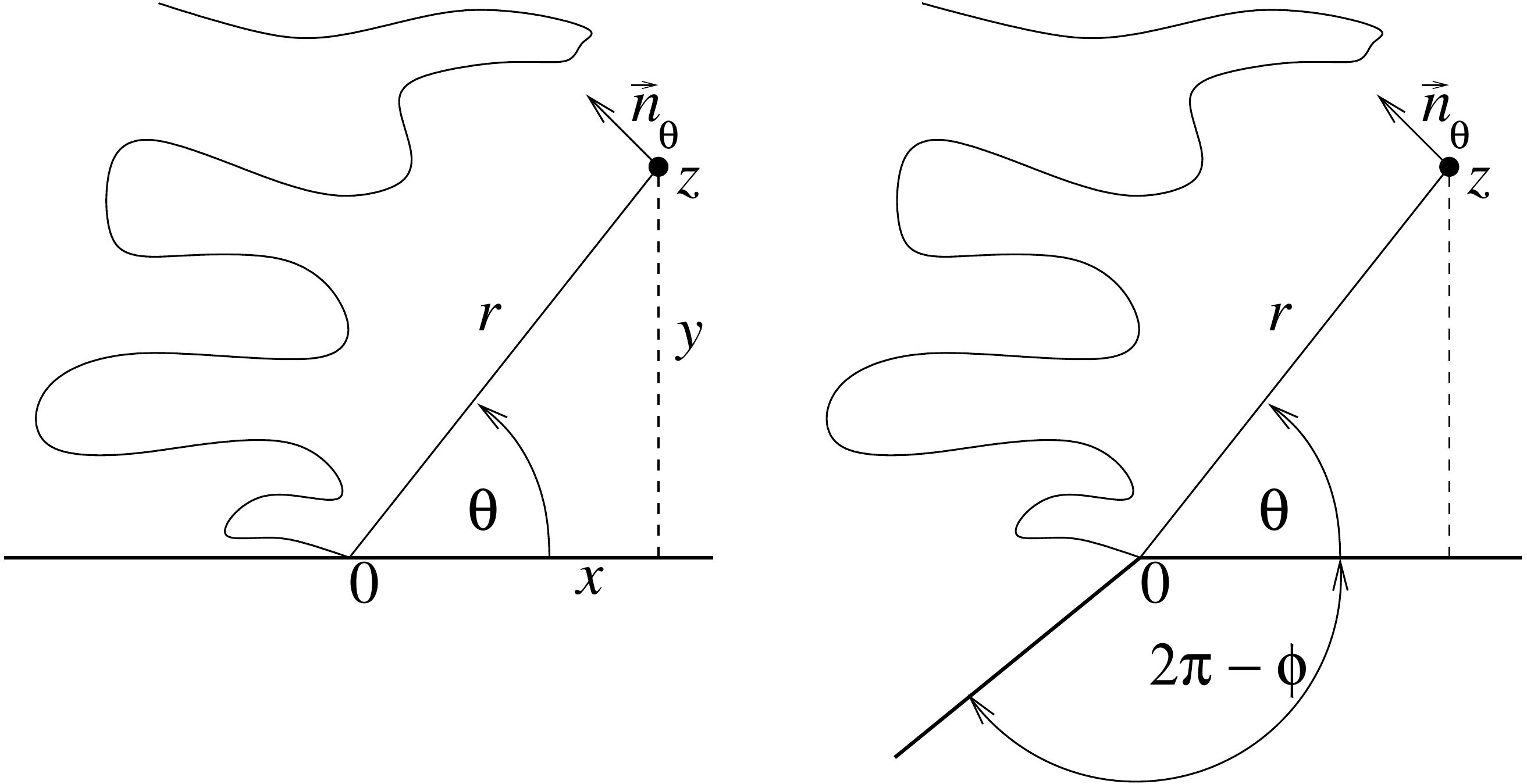}
%\leftline{\fbox{$z$}}\vspace*{-0.3cm}
\end{center}
\vspace*{-0.51cm}
\caption{Left: Geometry ${\cal A}$: A self-avoiding polymer fixed at the origin and constrained to remain 
left of the point $z$.
Right: Geometry ${\cal B}$: same as ${\cal A}$, the polymer being fixed at the top of a wedge of exterior angle $\phi$.}
\label{f:1}
\end{figure}%
\begin{figure}%[t]
\begin{center}
\leftline{\fbox{$z$}}\vspace*{-0.3cm}
\fig{0.65\columnwidth}{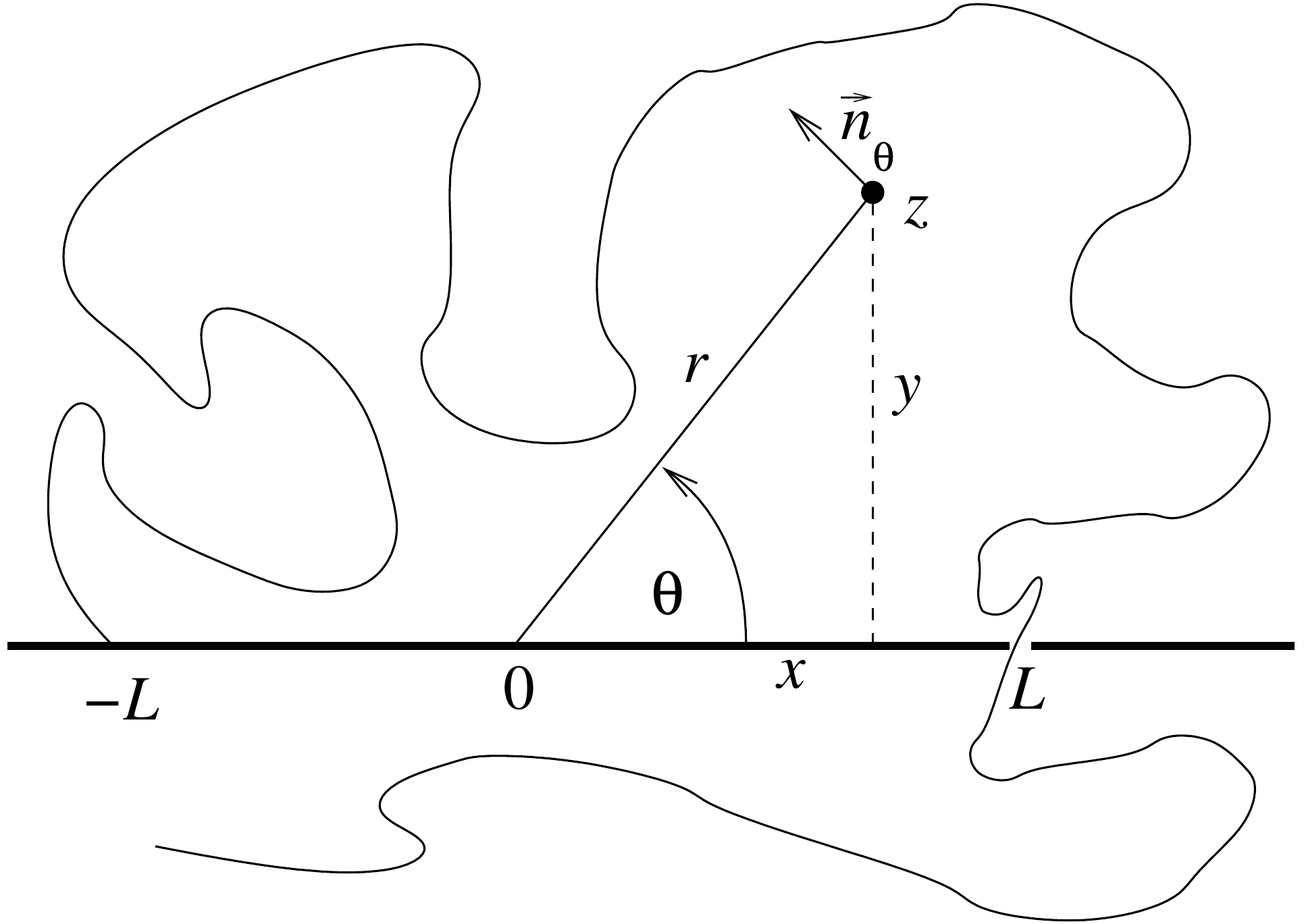}
\end{center}
\vspace*{-0.05cm}
\caption{Geometry ${\cal C}$: A self-avoiding polymer constrained to depart from $x=-L$, reaching $x=L$, and encircling both the origin and point $z$.}
\label{f:2}
\end{figure}
%
%\begin{figure}
%\begin{center}
%\leftline{\fbox{$w$}}\vspace*{-0.3cm}
%\fig{0.6 \columnwidth}{geom}
%\end{center}
%\caption{Geometry ${\cal A}$: A self-avoiding polymer fixed at the origin and constrained to remain 
%left of the point $z$.}
%\label{f:1}
%\end{figure}%
%\begin{figure}
%\begin{center}
%\leftline{\fbox{$z$}}\vspace*{-0.3cm}
%{\hspace{1cm}\fig{0.5\columnwidth}{geom4}}
%\end{center}
%\caption{Geometry ${\cal B}$: same as ${\cal A}$, the polymer being fixed at the top of a wedge of exterior angle $\phi$.}
%\label{f:4}
%\end{figure}%

While there are many results available in 2-dimensional critical systems, some recent originating from stochastic L\"owner evolution (SLE) \cite{schramm00} (see \cite{Cardy2005,sleBernardBauerRev06} for review),
their implications in terms of fluctuation-induced forces has to our knowledge not been discussed.

In this Letter, we consider a polymer restricted, e.g.\ by plates or through absorption \cite{ErcoliniValleAdamcikWitzMetzlerRiosRocaDietler2007}, to a planar geometry, modeled by a self-avoiding walk (SAW) of $N$ steps on a 2d lattice of spacing $a$. In the limit of large $N$ and small $a$ it is described by a continuum model. Start with a polymer with {\it one endpoint fixed}.  Geometry $\cal A$ represented on the left of Fig.\ \ref{f:1} is  a half plane where the polymer's end is fixed at the origin and free to wander to infinity. Then place a mesoscopic object, modeled by a disk of size $\ell$, at point $z=x+i y$. The object is impenetrable to the polymer, which is hence constrained to remain on the left of point $z$. We are interested in the free energy
\begin{eqnarray}
{\cal F} = - k T \ln {\cal P}(z,\bar z) \ .
\end{eqnarray}
${\cal P}(z,\bar z) = Z(z,\bar z)/Z$ where $Z$ is the partition sum of the polymer in the absence of the object and $Z(z,\bar z)$ is the constrained one. Since the SAW in the continuum limit is conjectured to be described by SLE with parameter $\kappa=8/3$ \cite{SchrammSAW,Kennedy},  we can use ${\cal P}(z,\bar z) ={\cal P}_0(\theta)$ as given by Schramm's formula (for $\kappa=8/3$) ${\cal P}_0(\theta) = \cos^2(\theta/2)$,
where $\theta$ is the angle with the real axis (see Fig.\ \ref{f:1}). From this we obtain the force exerted by the polymer on the impenetrable object:
\begin{equation}
 \vec f = - \vec \nabla F=  kT \frac{\vec n_\theta} r \frac{\partial }{\partial \theta} \ln {\cal P}_0(\theta) =  - kT \frac{\vec n_\theta} r \tan \left(\frac{\theta }{2}\right) \label{single}
\end{equation}
This result is valid in the (critical) limit $a ,\ell \ll r$, of object- and monomer-size  small compared to $r$. Note that when approaching  the boundary on the $x<0$ side, the object is repelled by a force diverging as $2 k T/y$, with $y$ the distance from the wall.

We can now use conformal invariance to obtain results in various geometries. The simplest one is the wedge geometry $\cal B$, see right of Fig.~\ref{f:1}, with exterior angle $\phi$, the polymer being attached at the top of the wedge. Under the map $w=g(z)=z^{\pi/\phi}$ the wedge geometry (in coordinate $z=x+i y$) is mapped back to the half plane (in coordinate $w$). The case $\phi=2 \pi$ corresponds to the full plane with impenetrable positive real axis. Conformal invariance   means that ${\cal P}(z,\bar z)={\cal P}_0(g(z),\overline{g(z)})$ where ${\cal P}_0$ is the upper-half plane result given above. We find for the free energy and force
\begin{eqnarray}
 {\cal F}_{\cal B} &=& - kT\left[ \ln ( 1 + \cos(\alpha \theta)) - \ln2\right] \\
 \vec f_{\cal B}&=&- k T \frac{\vec n_\theta}{r} \frac{\pi}{\phi} \tan(\pi \theta/2 \phi)\ .
\end{eqnarray}
Let us now study a polymer with {\it two endpoints fixed} as shown in  Fig.\ \ref{f:2} (geometry ${\cal C}$). Note that since SLE describes the continuum limit of the SAW with fixed endpoints but fluctuating number of steps $N$ at the critical chemical potential \cite{SchrammSAW,Kennedy}, a possible setting for an experiment is to consider the real axis as impenetrable, fix one endpoint at $x=-L$ and place a hole at $x=L$, through which the self-avoiding polymer passes. Note that  it is also possible to use two symmetric holes. Assuming equilibrium for an infinitely long polymer ensures that the  chemical potential is at its critical value. We can now use $w=g(z)= \frac{z+L}{L-z}$ which maps geometry $\cal C$ back to $\cal A$. It maps the half plane onto itself, preserves the real axis, and maps $z=-L$ to $w=0$ and $z=L$ to infinity, hence back to Fig.\ 1. %(in the variable $w$). 
Note that  the segment $[-L,L]$ is mapped to the real positive $w$ axis. Conformal invariance % of SLE 
yields %the free energy 
\begin{equation}
{\cal F}_{\cal C} = -k T \left[ \ln \left(  \frac{\epsilon(L^2-r^2)}{\sqrt{r^4-2 \cos (2 \theta ) r^2 L^2+L^4}}{+}1\right)-\ln 2\right] 
\end{equation}
with $\epsilon=1$ if the object is inside the area encircled by the polymer and $\epsilon=-1$ if it is outside. Computing the force one finds that for $\theta=\pi/2$ the force is radial $f_r=- 2\epsilon k T/[r (1+r^2/L^2)]$ and crosses over from $1/r$ to $L^2/r^3$ as $r$ increases, being attractive if the object is inside, repulsive if it is outside. 

\begin{figure}[t]
\setlength{\unitlength}{0.01\columnwidth}
%\fboxsep0mm
{\begin{picture}(100,41)
\put(0,38){\fbox{$z$}}
\put(0,0){\fig{0.44\columnwidth}{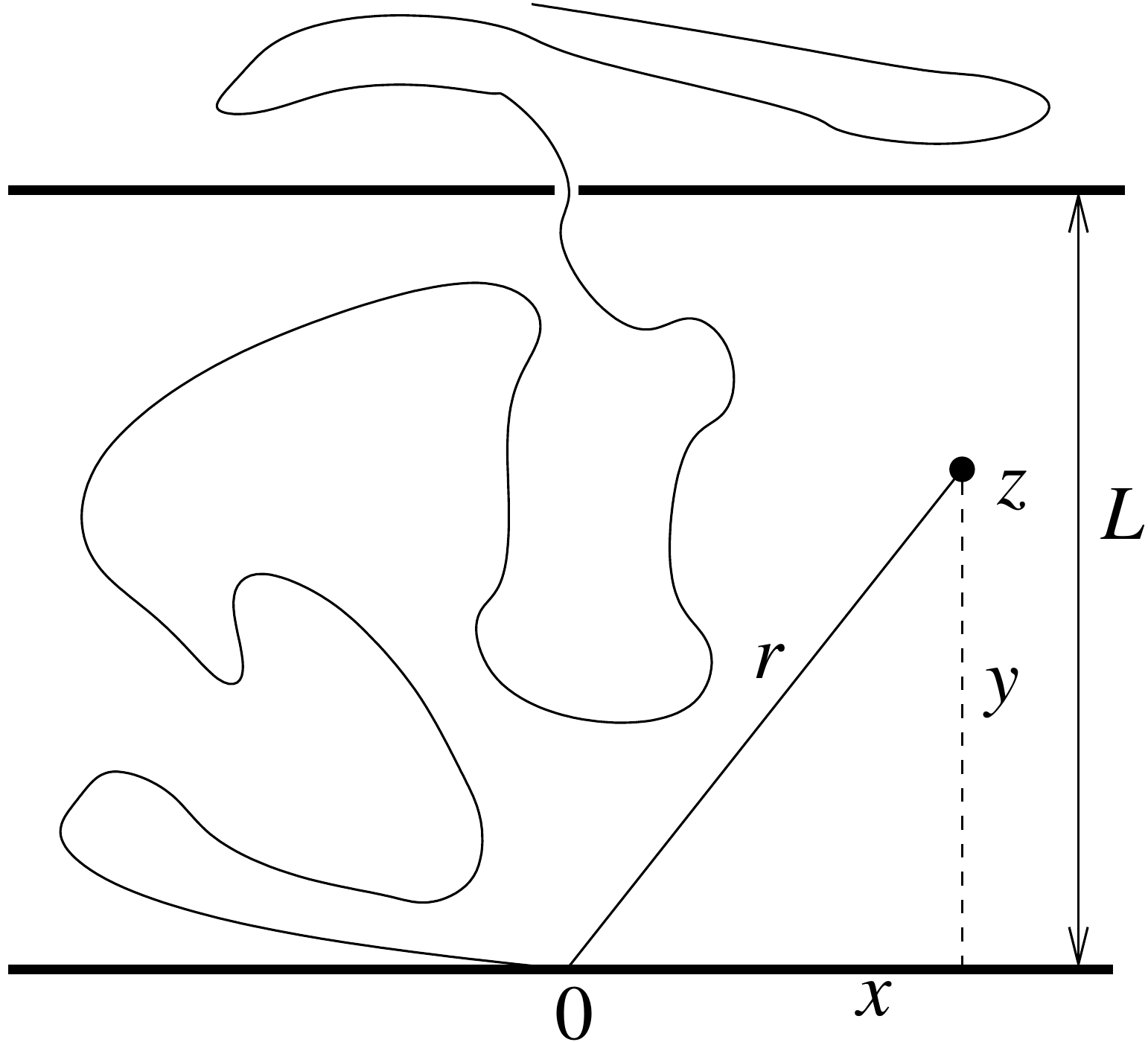}}
\put(48,30){\parbox{0.17\columnwidth}{\fig{0.17\columnwidth}{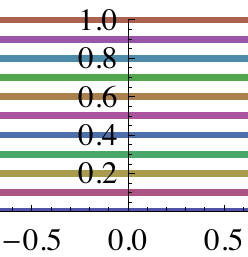}}$\to$\parbox{0.29\columnwidth}{\fig{0.29\columnwidth}{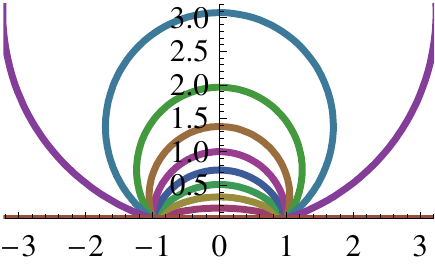}}}
\put(48,9){\parbox{0.17\columnwidth}{\fig{0.17\columnwidth}{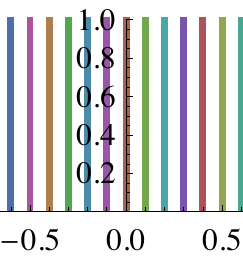}}$\to$\parbox{0.29\columnwidth}{\fig{0.29\columnwidth}{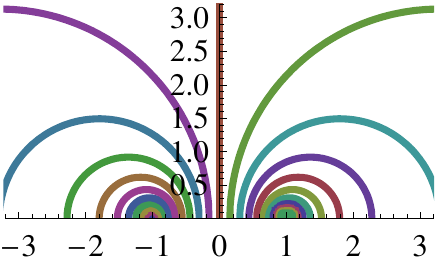}}}
\end{picture}}
%\leftline{\fbox{$z$}}\vspace*{-0.3cm}
%\fig{0.45\columnwidth}{geom3}
%\parbox{0.2\columnwidth}{\fig{0.2\columnwidth}{strip1}}$\to$\parbox{0.2\columnwidth}{\fig{0.2\columnwidth}{strip2}}\\
%\parbox{0.48\columnwidth}{\fig{0.48\columnwidth}{strip3}}$\to$\parbox{0.48\columnwidth}{\fig{0.48\columnwidth}{strip4}}
\caption{Left: Strip geometry ${\cal D}$: A self-avoiding polymer constrained to depart from $x=0$, passing through $x=i L$, and staying left of point $z$. Right: Mapping of the strip to the plane.}
\label{f:3}
\end{figure}

Instead of a half-plane one can compute the force in any singly connected domain, as e.g.\ a disk, or a strip. We consider two distinct infinite strip geometries $z=x+ i y$. In the first, $\cal D$, presented on Fig.\ \ref{f:3}, the strip is $0\le y\le L$ and the polymer is attached at $z=0$ and $z= i L$ (in the sense defined above, i.e.\ passing through a hole at $Z=iL $). Using $w=\tanh(\pi z/(2 L))=(e^{\pi z/L}-1)/(e^{\pi z/L}+1)$ to map it to  geometry $\cal A$ of Fig.\ \ref{f:1}, one finds the free energy in geometry $\cal D$:
\begin{equation}
{\cal F}_{\cal D} = - k T \ln \left[ \frac12 + \frac12 \frac{\sqrt 2 \sinh (\pi x/L)}{\sqrt{\cosh(2\pi x/L)-\cos (2\pi y/L)}} \right]
\end{equation}
On the symmetric line $y=L/2$ the force is directed along $x$ and equal to
$
f_x = \frac{k T}L  \frac{2 \pi }{1+e^{2 \pi  x/L}}, 
$ which has a finite limit at large negative $x$. 
\begin{figure}[t]
\vspace*{0.3cm}
\leftline{\fbox{$z$}}\vspace*{-0.6cm}
\begin{center}\fig{0.75\columnwidth}{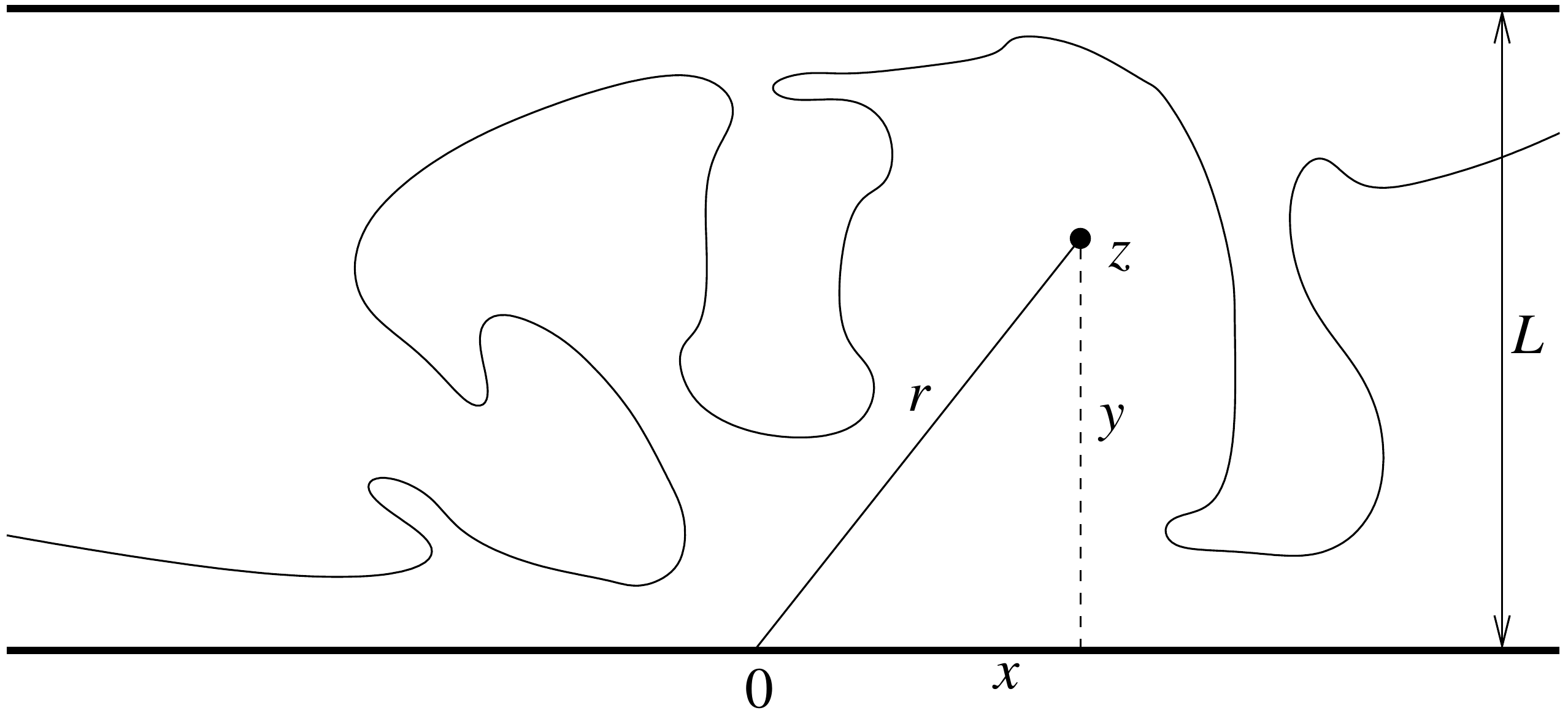}\end{center}
\parbox{0.16\columnwidth}{\fig{0.16\columnwidth}{strip5a}}$\to$\parbox{0.26\columnwidth}{\fig{0.26\columnwidth}{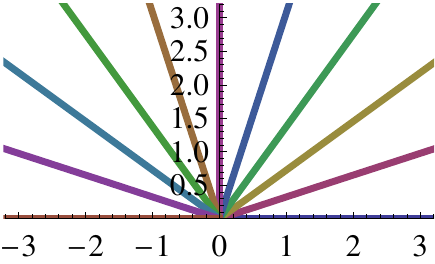}}\hspace{5mm}\parbox{0.16\columnwidth}
{\fig{0.16\columnwidth}{strip7a}}$\to$\parbox{0.26\columnwidth}{\fig{0.26\columnwidth}{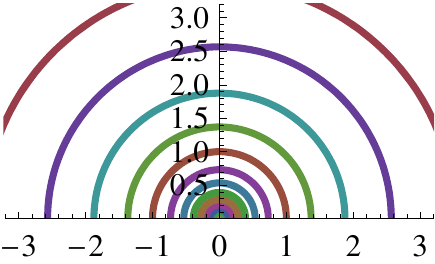}}
\caption{Top: Strip geometry ${\cal E}$: A self-avoiding polymer constrained to depart from $x=-\infty$, going to   $x=+\infty$, and passing at the top of point $z$. Bottom: Mapping of the strip to the plane.}
\label{f:5}
\end{figure}%

In the second strip geometry, $\cal E$ on figure \ref{f:5}, the polymer is attached infinitely far away on each side and the object is below it. Using the map $w=e^{\pi z/L}$, one finds the free energy and force (with $f_x=0$): 
\begin{eqnarray}
 {\cal F}_{\cal E} &=& - kT \left[\ln \left( 1 + \cos(\pi y/L)\right) - \ln2\right] \\
 f^{\cal E}_{y} &=& - \frac{\pi}{L} \tan\!\left({\pi y\over 2 L}\right)\ .  \label{strip2} 
\end{eqnarray}
In all cases considered above the force tends to bring the object towards a portion of the boundary. One can ask whether it is possible to levitate the object into a stable equilibrium away from the boundaries. For this one needs (at least) two polymers. This more difficult problem was solved when the two polymers start at the same point or nearby on the real axis and both go to infinity \cite{GamsaCardy2005}. We use their extension of Schramm's formula to two SLEs conditioned not to merge before reaching infinity. %(see also\cite{Burkhard} for earlier related results). 
%It is reasonable to conjecture that it applies to two mutually SAW and an impenetrable object at point $z=x+i y$, with $|z| \gg x_0$ (hence one may set $x_0=0$). 
One defines ${\cal P}_l$, ${\cal P}_m$ and ${\cal P}_r=1-{\cal P}_m-{\cal P}_l$ the relative weights of configurations such that the object is constrained to lie on the left of both polymers (l), in the middle (m) or to the right (r). Then ${\cal P}_m= \frac{4}{5} \sin^2(\theta)$, hence the free energy is:
\begin{equation}
 {\cal F}_m = - k T \left[ 2 \ln (\sin \theta) + \ln(4/5)\right]\ .
\end{equation}
More complicated formula hold for ${\cal P}_r$ and ${\cal F}_r$. 
%{\cal P}_r&=&\frac{2 t(13+15t^2)+(3 \pi + 6 \arctan(t)) (1+6 t^2+5 t^4)}{30 \pi (1+t^2)^2} \nonumber
We obtain for the force exerted on a point which remains to the left of the two polymers (l), in the middle (m) or to the right (r) as $\vec f= f^\theta \vec n_\theta$ with:
\begin{eqnarray}
f^\theta_{{l}} &=& -\frac{k T}{r} 
\frac{8 \sin (\theta ) [-12 \theta  \cos (\theta )+9 \sin (\theta )+\sin (3 \theta )]}{-24 \cos (2 \theta ) \theta -36 \theta +28 \sin (2 \theta
   )+\sin (4 \theta )} \nn \\
   f^\theta_{{m}} &=& \frac{k T}{r}  2 \cot (\theta )  \label{middlef} 
   \end{eqnarray}
\begin{figure}[t]
\setlength{\unitlength}{0.01\columnwidth}
%\fboxsep0mm
%\fbox
{\begin{picture}(100,35)
\put(0,0){\fig{0.48\columnwidth}{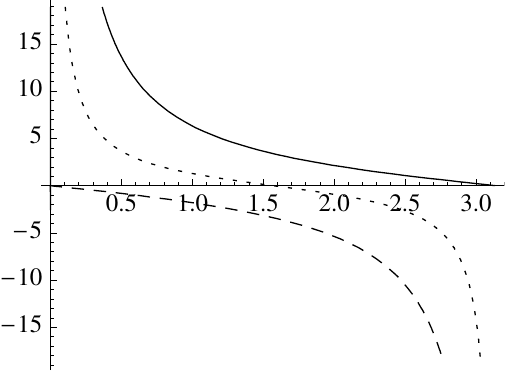}}
\put(52,0){\fig{0.48\columnwidth}{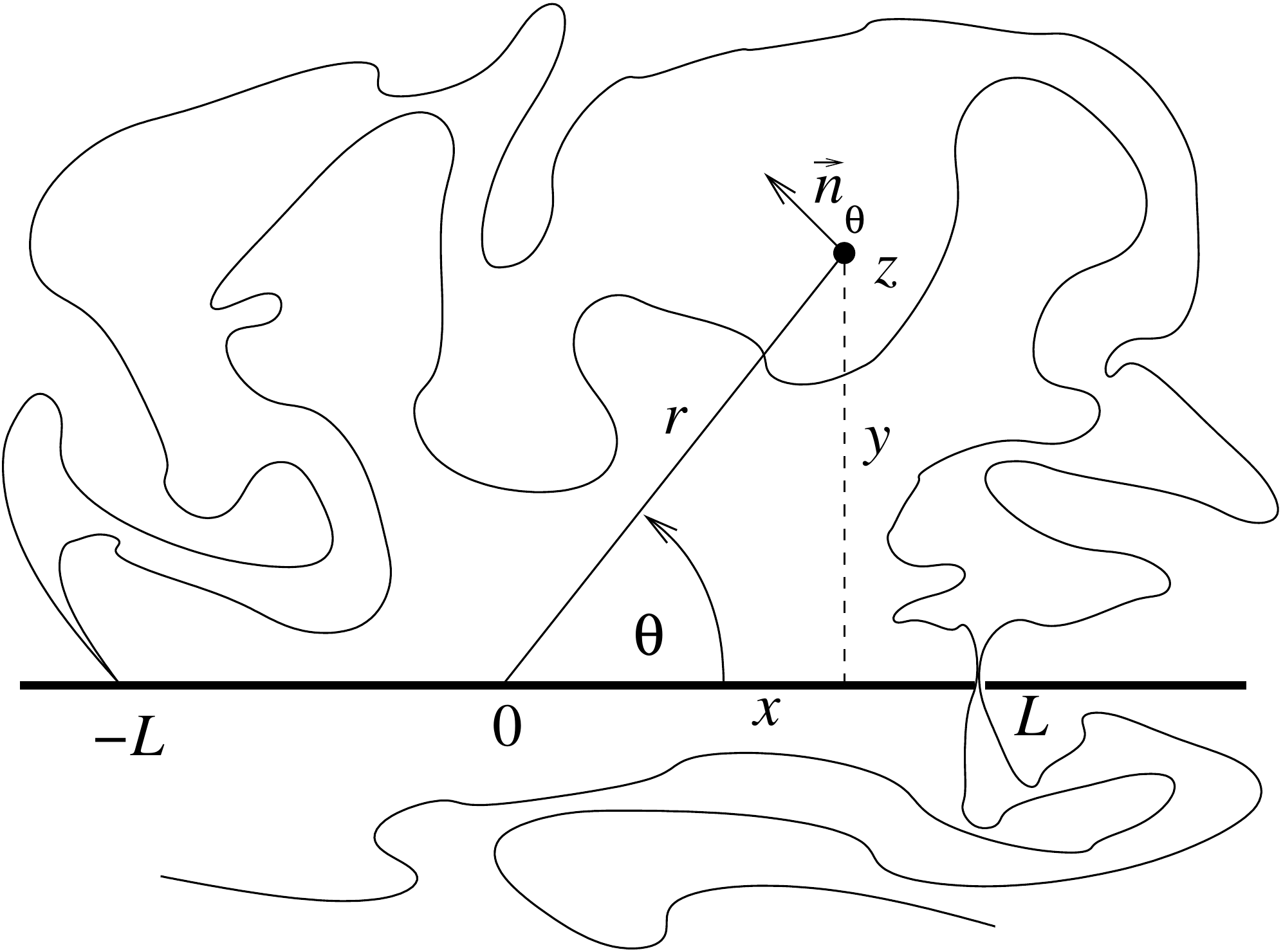}}
\put(46,18.5){$\theta$}
\put(0,32){$f$}
\end{picture}}
\fig{1 \columnwidth}{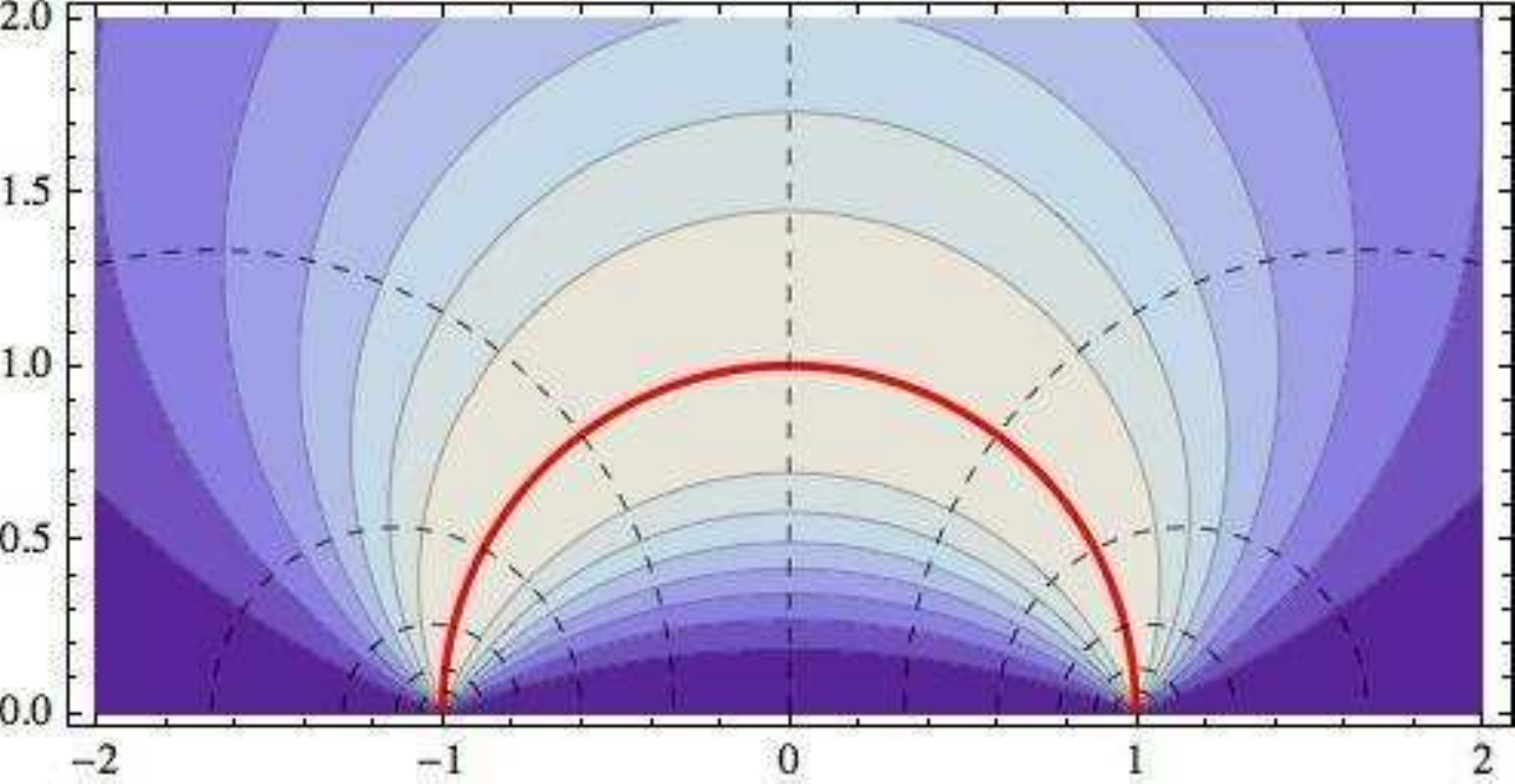}
\caption{Top left: Force along $\vec n_\theta$ exerted by 2 self-avoiding polymers on a point object, if the object is left of the two polymers (solid), between them (dashed) or right of them (dotted), in geometry ${\cal A}$. Top right: geometry ${\cal C}$. Two self-avoiding polymers fixed at $-L$, passing through a hole at $L$, and constrained to remain 
above and below the point  $z$. Bottom: Equal probability lines (solid) and force flow lines (dashed) for geometry ${\cal F}$; 
plot-units are $L$.}
\label{f:6}
\end{figure}%
%\begin{figure}[t]
%%\fig{0.7 \columnwidth}{geom2bis}\\
%%\smallskip
%\fig{1 \columnwidth}{2polycircle}
%\caption{}
%\label{f:2bis}
%\end{figure}%
and $ f^\theta_{\mathrm{r}}(\theta)= - f^\theta_{\mathrm{l}}(\pi-\theta)$. %, an obvious symmetry which also holds for a single polymer, where (\ref{single}) gives $\vec f_r$. 
This is plotted on Fig. \ref{f:6}. Note that when the object is trapped in the middle of the two polymers, the symmetry line $\theta=\pi/2$ is a line of equilibrium points, stable in the angular direction and neutral in the radial one. Hence the object is brought back to the symmetry line and force flow lines are circles $r=cst$ heading towards $\theta=\pi/2$. A remarkable property holds:
\begin{equation}\label{11}
{\cal P}_m(z,\bar z)= \frac{16}{5} {\cal P}^{(1)}_{l}(z,\bar z) {\cal P}^{(1)}_r(z,\bar z)\ ,
\end{equation}
where ${\cal P}^{(1)}_{l/r}$ is the (Schramm) probability for a single self-avoiding polymer to pass left/right of the point. Hence, if the point is in the middle, the fluctuation force is {\it the same} as for two independent polymers, i.e.\ mutual avoidance does not change the result, as can be checked on (\ref{middlef}). This is {\it not true} if the polymers are on the same side of the object. 

Let us consider again the geometry of Fig.\ \ref{f:2} with now both polymers attached at $x=-L$, and both passing through a hole at $x= L$, see top right of figure \ref{f:6}. An object trapped in the middle acquires a free energy:
\begin{equation}
{\cal F}_m=
-k T \log \left(\frac{16 r^2 L^2 \sin ^2(\theta )}{5 \left(r^4-2 \cos (2 \theta ) r^2L^2+L^4\right)}\right)\ .
\end{equation}
The equipotential lines are given on the bottom of figure \ref{f:6}, with the minimum on the circle of radius $L$, passing through $\pm L$ (bold red). 
This leads to a force
\begin{eqnarray}
\vec  f_{{m}} = \frac{2 kT}{r} \frac{(L^4-r^4) \vec n_r + (r^2-L^2)^2 \cot(\theta) \vec n_\theta}{L^4 + r^4 - 2 r^2 L^2 \cos(2 \theta)}\ .
\end{eqnarray}
which due to (\ref{11}) is the sum of the forces of two independent SAWs. There is now a semi-circle of equilibrium points $r=L$, which is the image of the vertical straight line passing through 0 of geometry $\cal A$. %. It is the image in $z=f(w)=\frac{w-1}{w+1}$, with $f$ inverse map of $g$, of the line $w$ imaginary of equilibrium points in geometry ${\cal A}$ with two polymers. 
Note that there is no force on the line, thus no  {\it stable} equilibrium.
%, yet: although the object is brought back to the equilibrium curve (flow lines are now circles in Fig.) , it can still drift along the neutral direction.

\begin{figure}[t]
\setlength{\unitlength}{0.01\columnwidth}
\fboxsep0mm
%\fbox
{\begin{picture}(100,35)
\put(0,0){\fig{0.317\columnwidth}{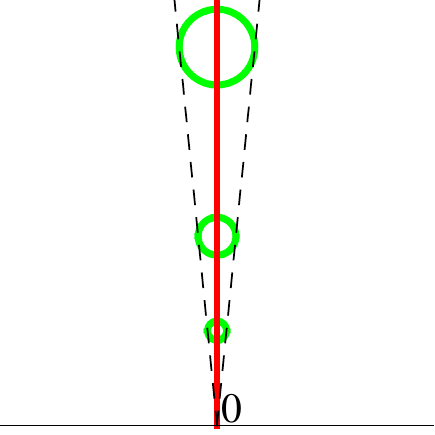}}
\put(32,17){$\to$}
\put(37.3,0){{\fig{0.627\columnwidth}{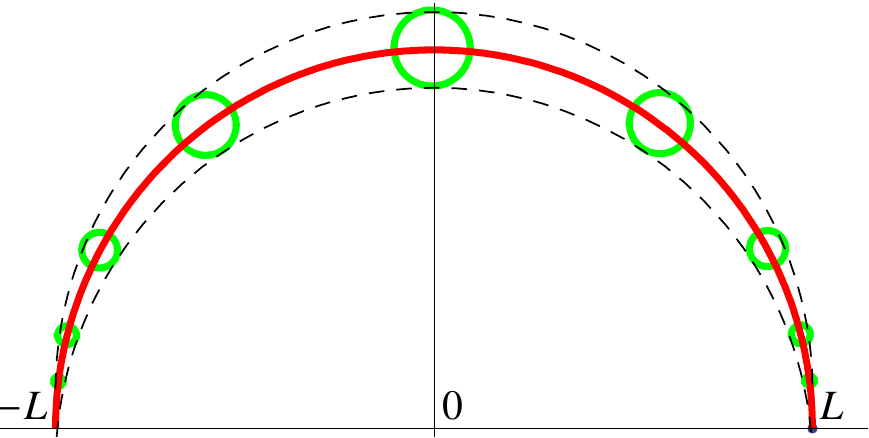}}}
\put(15,21){$\phi$}
\end{picture}}
\caption{Mapping of self-avoiding walks constrained to pass left and right of disks from geometry $\cal A$ to geometry $\cal C$: all disk drawn correspond to the same probability, i.e. same free energy}
\label{f:beau}
\end{figure}%
We now argue that trapping occurs in two cases (i) a finite-size object, e.g.\ a small disk; and (ii) a point submitted to a thermal bath. From scale invariance the probability ${\cal P}_m(\phi)$ that two SAWs starting at $0$ avoid a disk with center on the imaginary axis and pass one left, one right, depends only on the angle $\phi$ of the cone drawn on the left of Fig.\ \ref{f:beau} and is clearly
a decreasing function  of $\phi$, with ${\cal P}_m(0)={\cal P}_m$ and ${\cal P}_m(\pi)=0$. Hence a disk of fixed size will be pushed to infinity along the imaginary axis. Under conformal mapping of geometry $\cal A$ to $\cal C$, discs map to discs and the cone to the space between two circular arcs. Assuming conformal invariance of the probabilities, all disks shown in Fig. \ref{f:beau} have the same free energy. The center of a disk 
%with radius $\rho$ in geometry $\cal C$ 
will thus be pushed to the stable equilibrium point %$y= i \sqrt{L^2+\rho^2}$, 
above the origin, 
where the largest disk is drawn. %{\red **** K. didn't find simple way to prove this formula.***} 
A quantitative result is possible for small radius $\rho$. E.g.\ in the geometry ${\cal A}$ the no-hit probability for a disk centered at $x+i y$ reads $p \approx 1 - c (\frac{\rho}{2 y})^{2/3} \sin^2(\theta)$ to lowest order in powers of $\rho/y$, as extracted from \cite{beffara,bauer,Cardy2005}, with an unknown constant $c$. This gives the force $k T \vec \nabla \ln p$. %on a {\it penetrable} object with a high potential barrier. 
In the symmetric case $\theta=\pi/2$, 
%then ${\cal P}_{l} = \frac{1}{2} p$ and one obtains 
the force %on an inpenetrable disk 
along the radial direction is 
\begin{equation}\label{14}
f_r \approx \frac{2 c k T}{3} \frac{1}{r} \left(\frac{\rho}{2 r}\right)^{2/3} 
\end{equation}
which decays as $1/r^{5/3}$ at large distances. 

Consider now a point-like object %which can move along the force lines and 
subjected to the Casimir force above plus a thermal bath at temperature $T'$. The equilibrium Gibbs measure for the object is ${\cal P}={\cal P}(z,\bar z)^{T/T'}/{\cal Z}$, and the partition sum of polymer plus object is ${\cal Z} = \int \rmd^2 z\, {\cal P}(z,\bar z)^{T/T'}$. Here ${\cal P}$ is either Schramm's probability ${\cal P}_0$ for a single polymer, given above Eq.\ (\ref{single}), or ${\cal P}_m$ in Eq.\ (\ref{11}) for an object caught between two SAWs. For the latter case, equiprobability lines are plotted at the bottom of Fig.\ \ref{f:6} for $T'=T$. Depending on the geometry and $T/T'$, ${\cal Z}$ is either infinite, and the object diffuses to the region where the integral is divergent, or finite and the object is bound. The latter occurs for any $T'<T$ in geometry ${\cal C}$ (top right of Fig.\ \ref{f:6}), since at large $r$, ${\cal P} \rmd^2 z\, \sim \rmd\theta\, r \rmd r (\sin^2(\theta)/r^2)^{T/T'}$. For $T=T'$, a natural choice when the two polymers and the object are in mutual thermal equilibrium, this geometry is critical, hence the object diffuses to infinity. Other geometries however exhibit a bound state for $T=T'$. %Kay has ${\cal P}=\frac{16 r^2 \sin ^2(\theta )}{5 \left(r^4-2 \cos (2 \theta )
%   r^2+1\right)}$, thus  ${\cal P} \rmd r=\rmd r\, \frac85 \frac{\pi r}{1+ r^2}$ for $r>1$, and ${\cal P} \rmd r=\rmd r\, \frac85 \frac{\pi % r^3}{1+ r^2}$ for $r<1$.
E.g.\ the strip geometry ${\cal D}$ has a normalizable distribution, 
\begin{equation}
{\cal P}_m = \frac{\pi}{L^2 \ln 2} \frac{\sin^2(\pi y/L)}{\cosh(2 \pi x/L)-\cos(2 \pi y/L)} 
\end{equation}
and an exponentially localized bound state, with the length set by the strip width. An algebraic bound state is obtained if, in Fig.\ \ref{f:6} with the two polymers going through $-L$ and $L$, one rotates the real negative axis around $0$ clockwise to form a wedge with angle $\phi<\pi$. Then:
\begin{equation}
{\cal P}_m(r,\theta) = {\cal N}_a \frac{L^{2 a-2} r^{2 a}  \sin^2(a \theta)}{L^{4 a}+ r^{4 a} - 2 (r L)^{2 a} \cos(2 a \theta)}
\end{equation}
with $a=\pi/\phi>1$ (the formula remains true for $a<1$ as a non-normalizable density) and $\pi {\cal N}_a=4 a^2/(
\psi(\frac{1}{2 a})-\psi(\frac{1}{2}+\frac{1}{2 a})+a+ (\pi / \sin(\frac{\pi }{ a}))$. 
%Integration over the angle gives ${\cal P}_m(r)=\int_0^{\phi} r \rmd\theta  {\cal P}_m(r,\theta) = \frac{\pi  {\cal N}_a}{2 a}
% \frac{L^{-2} r^{2 a+1}}{r^{2 a}+L^{2 a}}$ for $r<1$, and $ {\cal P}_m(r)= \frac{\pi  {\cal N}_a}{2 a}  L^{-2} \frac{r}{r^{2 a}+1}$ for  $r>1$
%  which  has as ${\cal P}_m(r,\theta)$ its maximum at $r=(2 a-1)^{-1/(2 a)}$. 

Let us compare the force exerted by one and by two polymers. Let us choose the simplest geometry $\cal E$, the infinite strip with the two polymers attached at both ends (Fig.\ \ref{f:5}), where the force is along $y$. For an object in the middle, one has a restoring force towards the neutral axis  $y=L/2$ 
\begin{equation}
f^{\mathrm{m}}_y = %- \frac{2 kT \pi}{L} \tan\!\Big({\pi(y-\frac{L}{2})\over L}\Big)\ ,
 \frac{2 kT \pi}{L} \cot\!\Big({\pi y \over L}\Big)\ ,
\end{equation}
while  the force exerted by two polymers is $f_y =  k T \partial _y \ln {\cal P}$,  
\begin{eqnarray}
{\cal P} &=& \textstyle 24 \pi  \cos(\frac{2 \pi  y}{L})
  (1-\frac{y}{L})+36 \pi (1-\frac{y}{L})+28
   \sin(\frac{2 \pi  y}{L})\nn\\
   && \textstyle +\sin(\frac{4 \pi 
   y}{L}) \label{18}
\end{eqnarray}
Its ratio to the force  (\ref{strip2}) exerted by a single polymer increases monotonically from $\frac{16}5$ (at $y=0$) to $\frac{7}2$ (at $y=L$). For an interpretation of the first number see below.

We can now compute the force exerted by a single polymer on an object placed on the boundary of the system (e.g. the upper half plane $H$). We use the nice result of \cite{SchrammSAW} arising from the so-called restriction property obeyed by SAWs. It states that the probability that a SAW (from $0$ to infinity) does not visit a subdomain $A$ is $|g'_A(0)|^{5/8}$, where $g_A$ is the map  from $H\setminus A$ to $H$, which removes $A$ and has $g_A(0)=0$,
and $g_A(z) \sim z$ at infinity. Note that $H\setminus A$ must be singly connected, hence the object connected to the boundary. For a general domain $D$ and endpoints $a$ and $b$ on the boundary the probability is $|g'_A(a)|^{5/8} |g'_A(b)|^{5/8}$
with $g_A(a)=a$ and $g_B(b)=b$. Note that a similar result holds for a Brownian excursion, i.e. a Brownian from $a$ to $b$ conditioned not to hit the boundary, with the exponent $5/8$ replaced by $1$. Finally let us mention that for a
SAW from point $a$ on the boundary to point $b$ in the bulk (radial SLE) the probability becomes
$|g'_A(a)|^{5/8} |g'_A(b)|^{5/48}$. In CFT language $h_{1,2}$ (with $h_{1,2}=5/8$ for $\kappa=8/3$) is the dimension of the operator $\Phi_{12}$ creating a curve on the boundary, $2 h_{0,1/2}=5/48$ is the dimension of the bulk operator $\Phi_{0,1/2}$ creating
a curve in the bulk. $\Phi_{1,3}$ with $h_{1,3}=2$ creates two curves on the boundary conditioned not to annihilate. When generalized, this
implies that the force exerted by $n$ polymers with identical endpoints on a given subdomain $A$ connected to the boundary is proportional to $h_{1,n+1}=n(3 n+ 2)/8$, which explains the ratio $h_{1,3}/h_{1,2}=16/5$ found 
above, see Eq.\ (\ref{18}), for small $y$ (point close to the boundary)\footnote{For $n$ polymers ending in the bulk the
exponent $5/48$ is replaced by $2 h_{0,n/2}=\frac{3}{8} (\frac{n^2}{4} - \frac{1}{9})$.}.

The simplest example for an object $A$  connected to the boundary is a vertical segment $z=a+i y$ with $y \in [0,h]$, which is removed by the map $g_A(z) = \sqrt{(z-a)^2+h^2} + {\sign}(a) \sqrt{a^2+h^2}$. The no-hit probability is ${\cal P} = \big( \frac{a^2}{h^2+a^2}\big) ^{ 5/ {16}}$, and the total force $\vec f=k T \vec \nabla \ln {\cal P}$ is:
\begin{equation}\label{19}
f_x = \frac{5}{8} k T \frac{h^2}{a(a^2+h^2)}\ , \quad f_y = -  \frac{5}{8} k T  \frac{h}{(a^2+h^2)} \ .
\end{equation}
To obtain the force when the polymer starts at $0$ and ends at $z_0=x_0+i y_0$ in the half plane, one uses the map
$v=\tilde g_A(z)$ which preserves $z_0$ rather than $\infty$. Composing $g_A$ with a Moebius map which maps $H$ to $H$, $0$ to $0$ and $g(z_0)$ back to $z_0$, one finds a complicated formula which simplifies for
$x_0=a$ to  ${\cal P}= (\frac{a^2}{a^2 + h^2})^{\frac{5}{16}} y_0^{-5/12} (y_0^2-h^2)^{5/24}$. This gives for the force on the wall $f_y=-\frac{5k T }{24}(\frac{3h}{a^2+h^2}+\frac{2h}{y_0^2-h^2})$ which diverges as $y_0 \to h^+$.

Another example is a half disk of radius $r$ centered at $x=a>0$. The uniformizing map is $g(z)=z+\frac{r^2}{a}+\frac{r^2}{z-a}$. Hence the-no hit probability is ${\cal P}=(1 - \frac{r^2}{a^2})^{5/8}$, and the object is repelled with a force $f_x=\frac{5 kT}{4 a} r^2/(a^2-r^2)$.

%For the force along $x$, we find
%$
%f_x = \frac{5}{8} k T \frac{h^2 (y^2-a^2)}{a(a^2+h^2)(a^2+y^2)} 
%$, valid as long as $y$
%{\em *** Kay a enleve le terme en $a^4$. Even this result is strange, with the sign-change for $y=a$!}

The {\it polymer piston} is interesting for {\it extreme-value statistics}. Consider the strip geometry ${\cal D}$ on Fig.\ \ref{f:3} and add an impenetrable region ${\sf P}$ (the piston) for $x>a$. The map $h_a(z)= [\cosh(\frac{\pi}{L}a)-\cosh(\frac{\pi}{L}(z-a))]/[\cosh(\frac{\pi}{L}(z-a))+\cosh(\frac{\pi}{L}a)]$ maps the strip minus ${\sf P}$ to the upper half plane, and both axes $y=0$ and $y=L$ to $0$. Hence the map which removes the piston is $g_A(z)=h_{\infty}^{-1}(h_a(z)) =(L/\pi) \ln(\cosh(\pi(z-a)/L)/\cosh(a \pi/L)) $ while leaving $0$ and $i L$ fixed. The no-hit probability  is
\begin{equation}
 {\cal P} = |g'_A(0)|^{5/8}  |g'_A( i L )|^{5/8} = [\tanh(a \pi/L)]^{5/4}
\end{equation}
Note that this is also the cumulative distribution of $x_{\max}=a$, the {\it maximum excursion of a SAW}. The total force exerted on the piston is $f_x = 5 \pi/[2 L \sinh(2 a \pi/L)]$.

\begin{figure}[t]
\leftline{{\fig{\columnwidth}{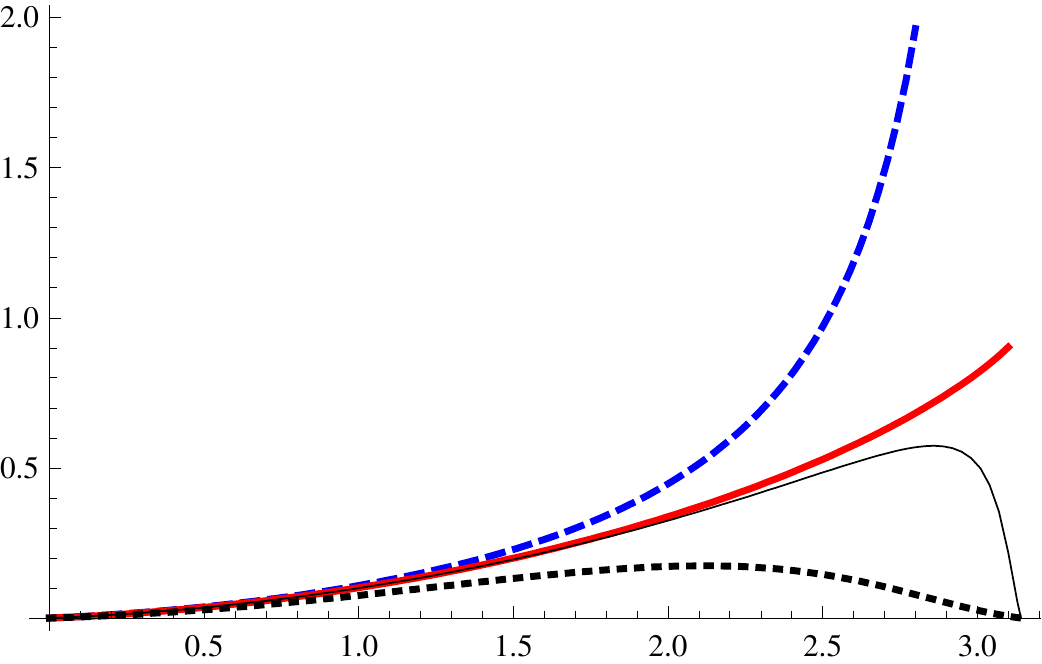}}\hspace*{-0.85\columnwidth}\raisebox{25mm}[0mm][0mm]{\fig{0.55\columnwidth}{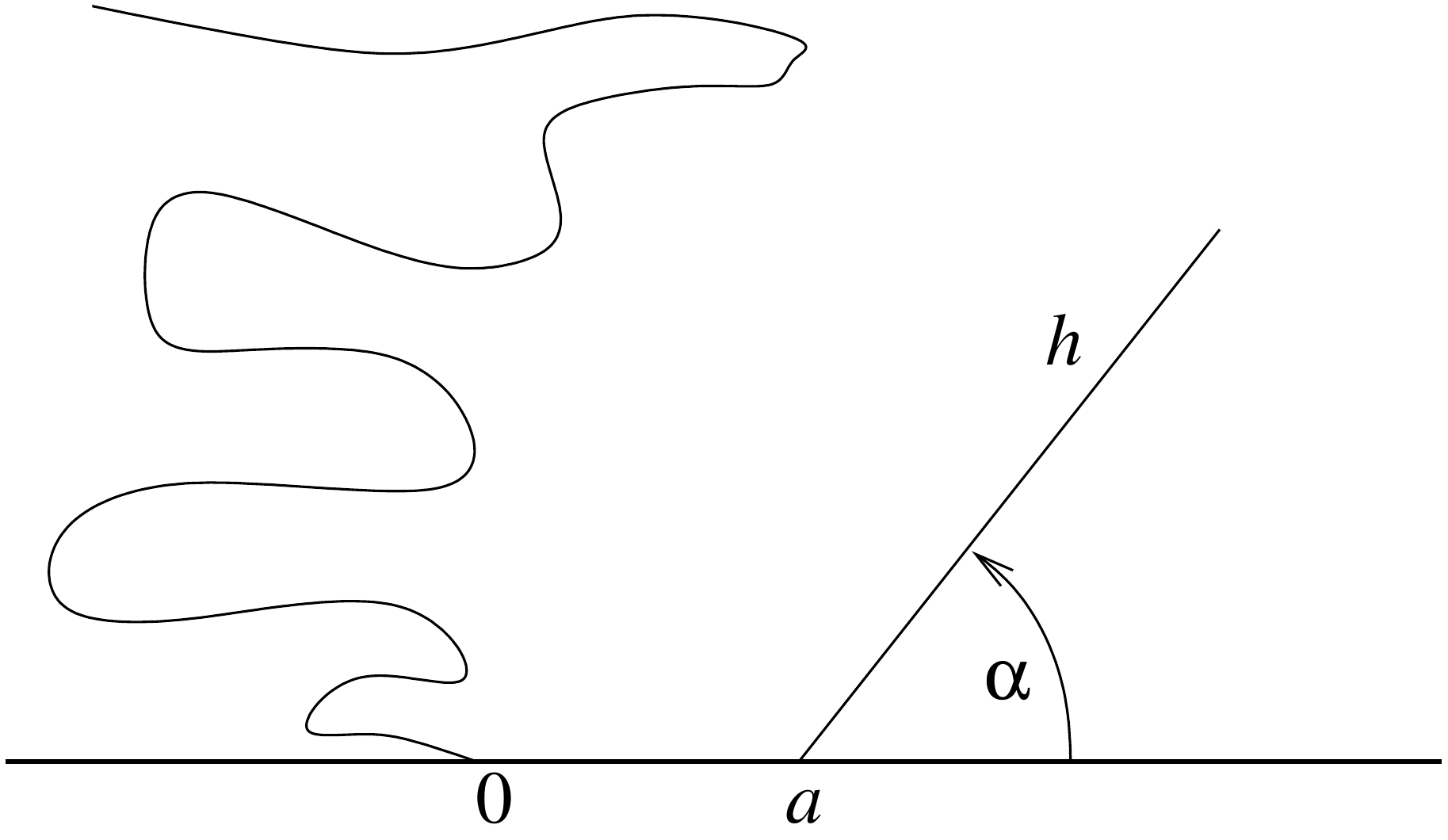}}}
\caption{Inset: Closing-door geometry. Main plot: $h/a$ times the free energy, i.e.\ $\frac ha \cal F$ (in units of $kT$) for this geometry, as a function of $\alpha$. The curves are (from top to bottom):  $h/a=2$ (dashed, blue), $h/a=1 $ (thick, solid, red), $h/a=0.95$ (thin, solid, black), $h/a=1/2$ (dotted, thick, black). The last point on $h/a=1$ curve is obtained analytically in (\ref{magic}).}
\label{f:door}
\end{figure}
\begin{figure}
{\Fig{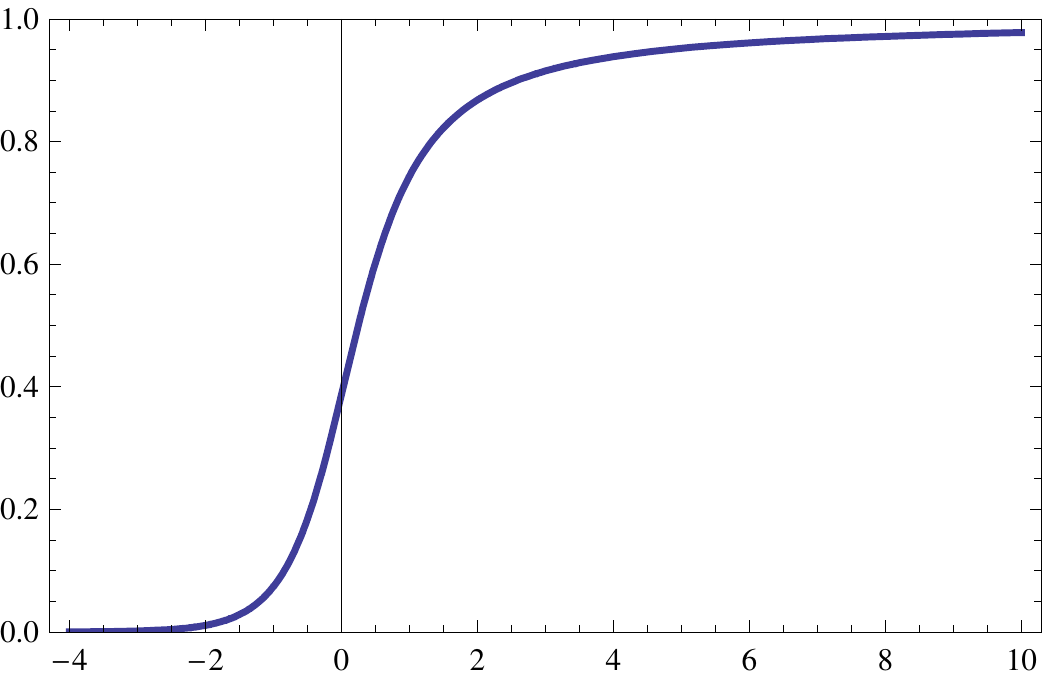}\hspace{-5cm}\parbox{0cm}{\raisebox{0.7cm}[0mm]{\fig{4.7cm}{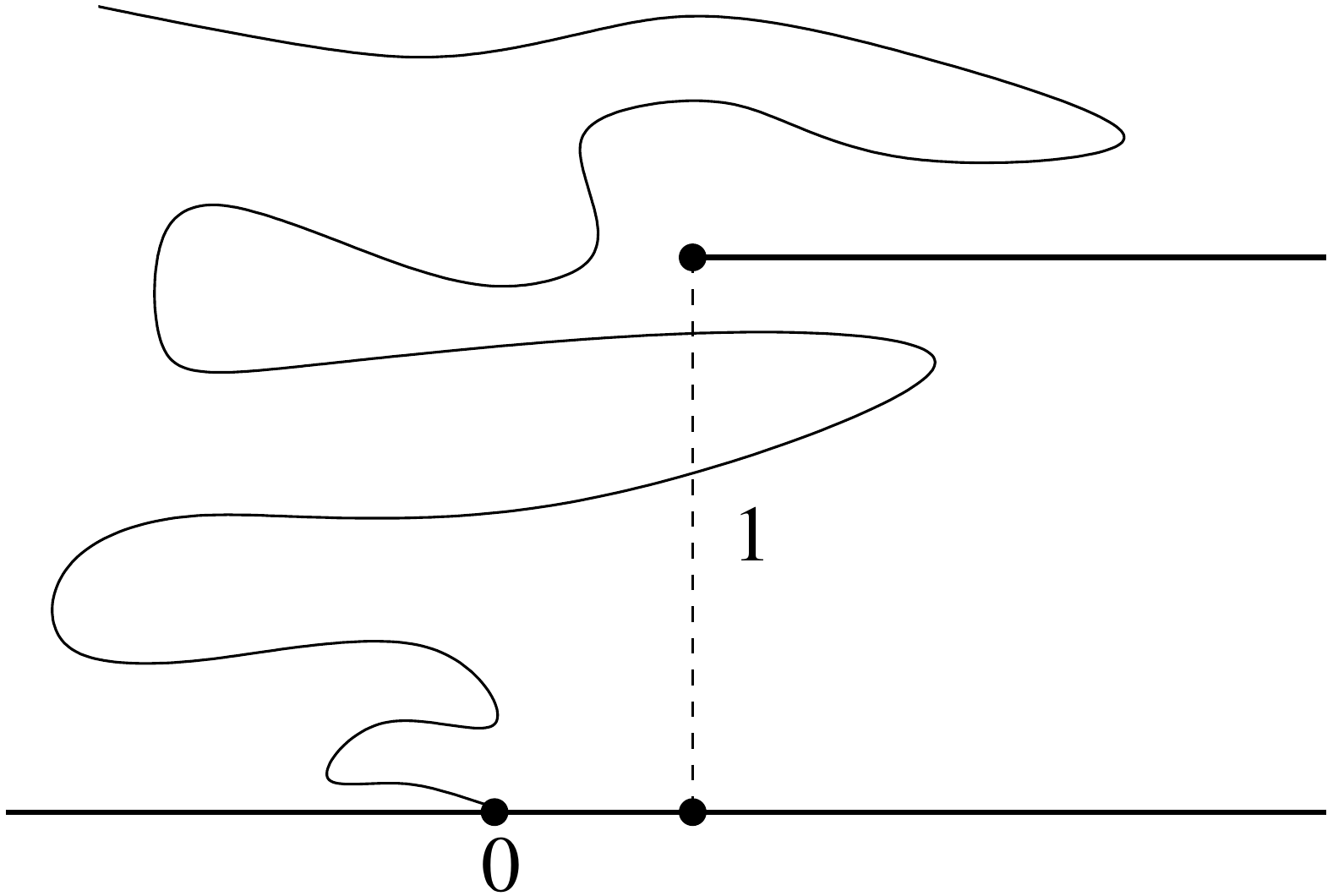}}}\hspace{2.1cm}\parbox{0mm}{\raisebox{0.73cm}[0mm]{$\ell$}}\hspace{2.9cm}}
\caption{The probability to avoid a wall starting from $(\ell,1)$ to $(\infty,1)$. Inset: the geometry in question.}
\label{f:N1}
\end{figure}

\begin{figure}[t]
\Fig{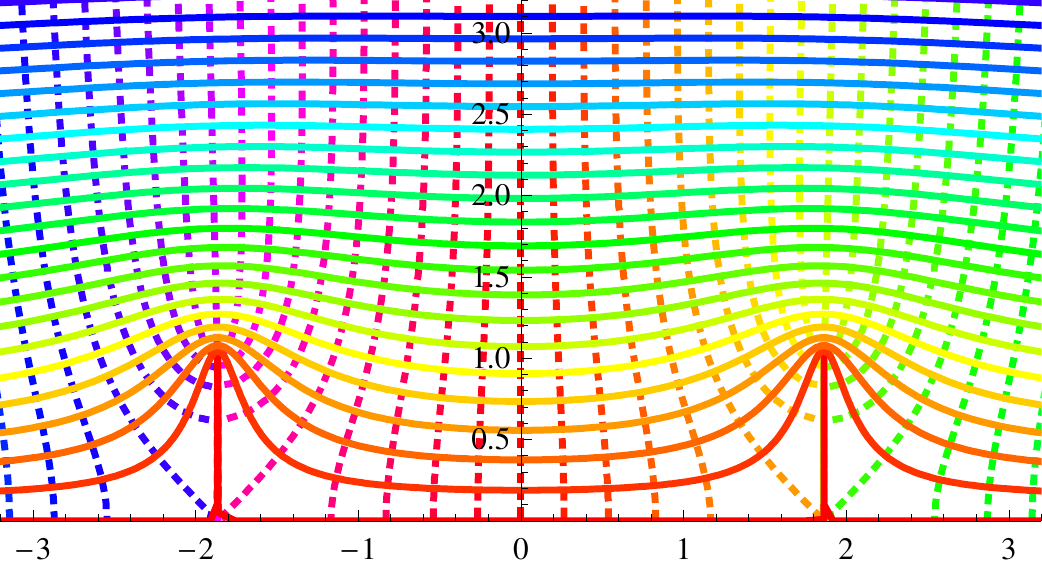}
\caption{Image of the upper half plane, and of lines parallel to the real axis (in thick), resp. imaginary (dotted), under the map $f(w)$ discussed in the text, which creates two slits, with $x_1=1$, $x_2=1.991$, and $x_3$=3. }
\label{f:2slits}
\end{figure}

\begin{figure}[t]
\fig{0.49\columnwidth}{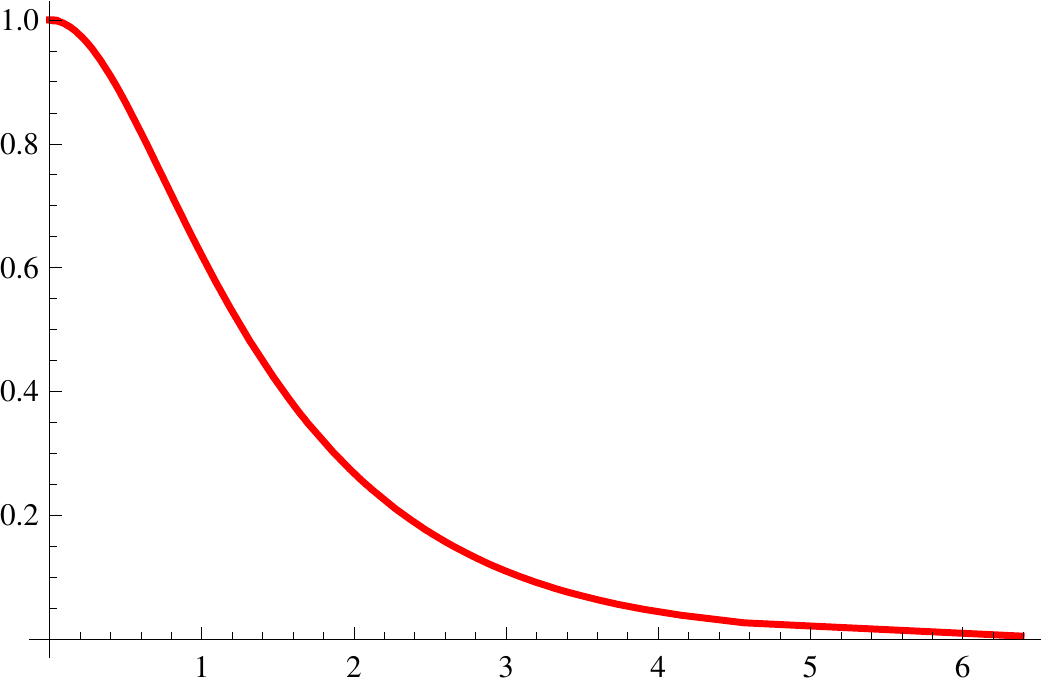} \fig{0.49\columnwidth}{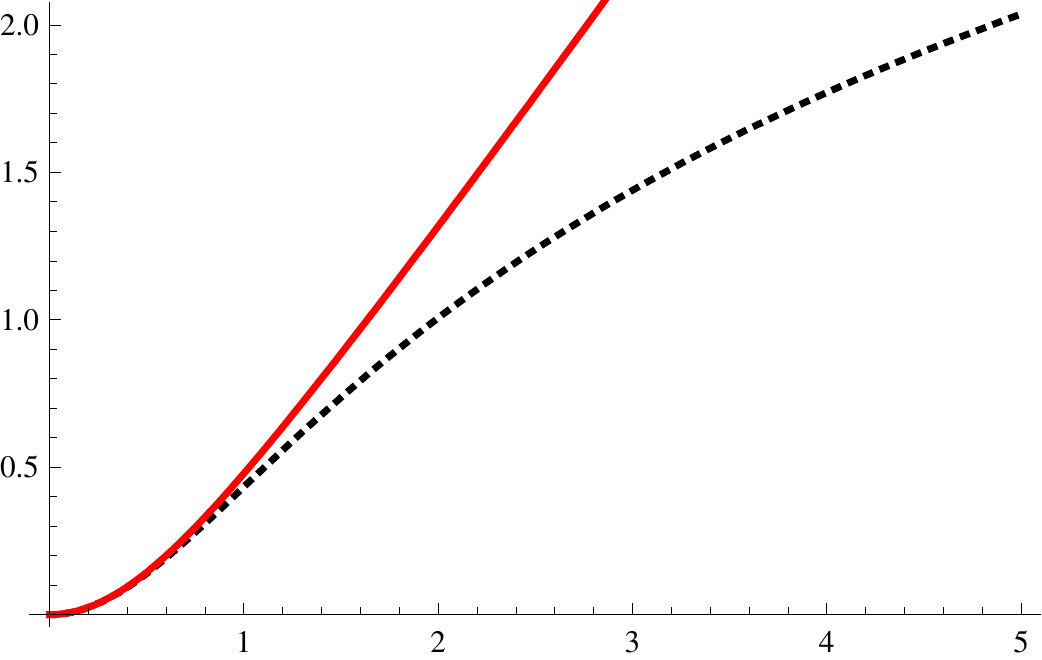}
\caption{Left: The probability that a polymer does not touch two slits, as a function of $h/a$ for the geometry of figure \ref {f:2slits}. Right: The free energy in units of $kT$ (solid line), compared to the sum of the free energies for a slit left and a slit right (dashed line), as a function of $h/a$. The difference is the interaction free energy mediated by the polymer.}
\label{f:2slitsprobaN}
\end{figure}

%
%\begin{figure}[t]
%\fig{0.7\columnwidth}{geom6}
%\caption{The door geometry.}
%\label{f:doorgeom}
%\end{figure}%
Consider now the ``door'' geometry, i.e. a segment $z=a+ t e^{i \alpha}$ with $t \in [0,h]$, of angle $\alpha=b \pi$. The relevant map $w=g(z)$ has an explicit form in terms of its inverse map $z=f(w)$ with 
$f(w)=a+(w-x_1)[(w-x_3)/(w-x_1)]^b$, $0<x_1<x_3$ with $a=x_1(x_3/x_1)^b$ and $h=b^b(1-b)^{1-b}(x_3-x_1)$. The no-hit probability is
\begin{eqnarray}\label{21}
 {\cal P} &=&\big[ \mu^{b} (1-b(1-\mu^{-1})) \big]^{-5/8} \\
 h/a &=& b^b(1-b)^{1-b} \mu^{-b} (\mu-1) \ ,
 \label{22}
\end{eqnarray}
where $\mu=x_3/x_1>1$ is solution of Eq.~(\ref{22}). The numerical solution is given on figure \ref{f:door}. An interesting limit
is represented in Fig.\ \ref{f:N1}, where $h=1/\sin(\pi k)$, $a=\ell + \cot(\pi k)$ and $k=1-b$ tends to zero. One finds that 
$\mu=\frac{1}{w k}  + O(k^0)$ with $w=W(e^{\ell \pi-1})$ the product-log function 
$W(z)$ solution of $z= W e^W$. This gives the no-hit probability of the horizontal half-line $i + x$ with $x>\ell$,
plotted on Fig.\ \ref{f:N1},
\begin{eqnarray}\label{magic}
{\cal P} &=& \Big[1+\frac{1}{W\left(e^{\ell\pi - 1}\right)}\Big]^{-5/8}\ .\qquad
%\\  &\to& 0.952578 \quad \mbox{for}\quad \ell=0\\
%\frac{\cal F}{kT} &=&  \frac{5}{8} \log \left(1+\frac{1}{W\left(\exp(\ell\pi -1 )\right)}\right)\nn
%\\  &\to& 0.952578 \quad \mbox{for}\quad \ell=0
\end{eqnarray}
Let us now consider the fluctuation force between two objects, here two identical slits, {\it mediated by the polymer}, here
in the symmetric position (see figure \ref {f:2slits}). 
Following \cite{SchwarzChristophel}, the map which produces two slits  is for $x_1<x_2<x_3$: 
$
f'(w) =  \frac{w^2-x_2^2}{\sqrt{w^2-x_1^2} \sqrt{w^2-x_3^2}}$, 
$   f(w)={ \textstyle  E\left(\arcsin ({\frac{w}{x_1})|\frac{x_1^2}{x_3^2}}\right) x_3}
+{F\left(\arcsin(\frac{w}{x_1})|\frac{x_1^2}{x_3^2}\right)(x_2^2-x_3^2)}/{x_3}$, 
where $E$, $F$ and $K$ (below) are the elliptic $E$, $F$ and $K$ functions, and our choice is $g(0)=0=f(0)$. 
The condition that $f(x_1)=f(x_3)$, or equivalently that $\Im f(x_3)=0$ yields a non-trivial condition. Define 
$\alpha:= x_1/x_3$, $\beta:=x_2/x_3$. Then for $0 < \alpha < \beta< 1$: 
$ 
\beta(\alpha)=\sqrt{\frac{E\left(\alpha ^2\right)-E\left(\arcsin \left(\frac{1}{\alpha }\right)|\alpha
   ^2\right)}{F\left(\arcsin \left(\frac{1}{\alpha }\right)|\alpha ^2\right)-K\left(\alpha
   ^2\right)}+1}
$.
The walls have position $\pm a$ and height $h$ (see figure \ref {f:2slits}):
\begin{eqnarray}
a=f(x_1)&=& \left(E\left(\alpha ^2\right)+\left(\beta ^2-1\right) K\left(\alpha
   ^2\right)\right)x_3 \\
 {h}=\Im {f(x_2)} &=&\Im \Big[ \textstyle E\left(\arcsin(\frac{\beta }{\alpha })|\alpha
   ^2\right) \nn\\&& \textstyle +\left(\beta ^2-1\right) \textstyle F\left(\arcsin(\frac{\beta }{\alpha
   })|\alpha ^2\right)\Big] x_3
\end{eqnarray}
The probability is 
$
{\cal P} = |f'(0)|^{-\frac 5 8} = |\frac \alpha{\beta^2}|^ {\frac 5 8}
$.
%with
%\begin{equation}
%|f'(0)| = \frac{x_2^2}{x_1 x_3}= \frac{\beta^2}\alpha
%\end{equation}
Fig.\ \ref {f:2slitsprobaN} shows a parametric plot of  ${\cal P}$, and of the interaction energy, as function of $h/a$.

Consider now a small smooth object described by $z=x+ i y$, $0 < y \leq Y (x)$, away from the origin, i.e. $Y(0)=Y(\infty)=0$. If we find a function
$f(t)$ with only positive fourier components $f_k$, such that $x=x(t)=\Re f(t)$, $Y(x)=\Im f(t)$ describes the boundary
for $t$ real, then $f(z)=z+\int_{k>0} f_k e^{i k z}=z+{ \frac1\pi}\int_t \frac{Y(x(t))}{t-z} $ is the inverse uniformizing map. In an 
expansion in powers of $Y(x)$ {\it and its derivatives} one finds $f(z)=\tilde f(z)-\tilde f(0)$ with $\tilde f(z)=
z + {\frac1\pi}\int_t \frac{Y(t)}{t-z} - \frac{1}{{2} \pi^2} \int_{t,t'} Y'(t) Y'(t') (\frac{1}{t-z} + \frac{1}{t'-z}) \ln|t-t'| +...$. This yields
the free energy
\begin{equation} \label{expansion}
{\cal F} = k T \frac{5}{8 \pi} \Big[ \int_t \frac{Y(t)}{t^2}  - \int_{t,t'} {\cal G}(t,t') Y'(t) Y'(t') +\ldots \Big]\ ,
\end{equation}
where ${ 2} \pi {\cal G}(t,t')=(t^{-2}+t^{\prime -2}) \ln|t-t'| + { 1/(tt')}$. For a single object 
centered at position $a$, $Y(t)=h(t-a)$, the repulsive force $f_x=-\partial_a \cal F$ decays as $f_x \approx 5 k T A/(4 \pi a^3)$
at large distances, %\footnote{$h<0$ leads to attraction since the polymer has a larger area to visit}
with a prefactor $A=\int_t Y(x(t))=\int_t h(t) - \frac{1}{\pi} \int_{tt'} h'(t)h'(t') \ln|t-t'| +O(h^3)$. In the case of two objects, (\ref{expansion}) yields their interaction, to lowest order, mediated by the
polymer. For small objects one finds ${\cal F}_{\mathrm{int}}= - k T {\frac{5}{4 \pi}} \partial_{a}\partial_b {\cal G}(a,b) \int_t h_a(t) \int_{t'} h_b(t')$. %with
%$- {  \pi} \partial_{a}\partial_b {\cal G}(a,b)=[{ a^4-2 b a^3+b^2 a^2-2 b^3 a+b^4}]/[a^3 b^3 (a-b)^2]$. 
%**** il reste le probleme de l'interpretation et de la validite **** 

The interaction of a small object at $z$ in the bulk with an arbitrary object on the boundary removed by the map $g(z)$
is obtained from the left passage probability $\cal P$, generalizing Schramm's formula to ${\cal P} = |g'(0)|^{\frac58}\frac12 \big[ 1+ \frac{\Re g(z)}{|g(z)|} \big]$.

The previous calculations can be extended to fluctuation forces for an object impenetrable to the interface described by SLE for any  $\kappa$. 
For illustration, the force in geometry $\cal A$ at $\theta = \pi/2$ reads
\begin{equation}
\vec f = 
 - k T \frac{\vec n_\theta} r
\frac{2 \Gamma \left(\frac{4}{\kappa }\right)}{\sqrt{\pi } \Gamma \left(\frac{4}{\kappa
   }-\frac{1}{2}\right)}\ .
\end{equation}
Extension to Ising at $T_c$ assumes that the object interacts only with the interface induced by changes in boundary conditions, not the bubbles proliferating at criticality, which seems artificial. Physically meaningful is the polymer at the $\Theta$ point \cite{DuplantierSaleur1987},  conjectured to  correspond to $\kappa=6$. Further results follow from recent works: (i) from \cite{hagendorf} one obtains the force exerted by a loop-erased random walk ($\kappa=2$) on an object of arbitrary shape. (ii) from the {\it double} left-passage probability of a SAW \cite{SimmonsCardy} around points $z_1,z_2$ one computes the Casimir interaction between two points. Interestingly, when they are close and away from the boundary the interaction force is {\it attractive} and diverges for $y\approx y_1\approx y_2$, $\theta\approx \theta_1\approx \theta_2$ as $|f| \sim k T  A (1-\cos \theta) y^{-2/3} |z_1-z_2|^{-1/3}$ with $A=- \sqrt{3 \pi} \Gamma(5/6)/(3 \Gamma(-2/3))=0.287457...$. 
Near the boundary for small $y_1=y_2=y$, ${\cal F}_{\mathrm{int}} = -k T y^4/(5 x_1 x_2 (x_1-x_2)^2)+ O(y^6)$, a {\it repulsive} interaction for $x_1<x_2/3$.

We thank M. Bauer, D. Bernard, T. Emig, C. Hagendorf, Y. Kantor and M. Kardar for useful discussions. This work was supported by ANR under program 05-BLAN-0099-01, and in part through NSF grant PHY05-51164 during the program Fluctuate08 at KITP.

\newpage

%\bibliography{citation}

%\newpage

\section{Appendix A: Radial SLE} 
Here we give details about a polymer starting at 0, and ending at $z_0$, and the force it exerts on the slit with height $h$ above $a$. 
We use the formula above Eq.\ (\ref{19}) that removes the slit, $g_A(z) = \sqrt{(z-a)^2+h^2} + {\sign}(a) \sqrt{a^2+h^2}$.
Note that the cut of the square root must be such that $g(z) \sim z$ at infinity, hence $g(x)<0$ for $x<a$, and $g(x)>0$ for $x>a$. Thus the cut is the real positive axis with $g_A(x+i \epsilon)=- \sqrt{(x-a)^2+h^2}$ for $x<a$. In other words, one defines in the full plane minus the real positive axis $\sqrt{x+i y}:=\frac{1}{\sqrt{2}}\left(  {\sign}(y) \sqrt{\sqrt{x^2+y^2}+x} + i \sqrt{\sqrt{x^2+y^2}-x}\right)$.

To obtain the force when the polymer starts at $0$ and ends at $z_0=x_0+i y_0$ in the half plane, one has to use the map
$w=\tilde g_A(z)$ which preserves $z_0$ rather than $\infty$. It can be constructed, by composing $g_A$ with a second Moebius  map. The latter maps $H$ to $H$,  $0$ to $0$ and $g(z_0)$ back to $z_0$. We claim that $w$ and $z$ are related by
\begin{equation}
\frac{1}{w} - \Re \frac{1}{z_0} = \frac{\Im(1/z_0)}{\Im(1/g(z_0))} \left[\frac 1{g(z)} - \Re \frac{1}{g(z_0)}\right]\ ,
\end{equation}
where $\Re$ and $\Im$ indicate real and imaginary parts. First, all $\Re$ and $\Im$ appearing above are real numbers. Thus, $w$ is a Moebius transform of $g(z)$ with real parameters. Second, $z=0$, i.e.\ $g(z)=0$ is mapped onto $w=0$. Third, $z_0$ is mapped to $w=z_0$. 
Using this, we find 
\begin{eqnarray}
\left| \tilde g_A'(0)\right| &=&\left| \frac{a z_0^2 \,\Im g_A(z_0)}{ \sqrt{a^2+h^2} y_0 g_A(z_0)^2} \right| \\
\left| \tilde g_A'(z_0)\right| &=& \left|\frac{y_0 g_A'(z_0)}{\Im  g_A(z_0)}\right|
\end{eqnarray}
This leads to 
\begin{eqnarray}
{\cal P}=\left|\frac{a}{\sqrt{a^2 + h^2}} \frac{z_0^2}{g_A(z_0)^2}\right|^{\frac{5}{8}} \left| \frac{\Im g_A(z_0)}{\Im(z_0)} \right|^{\frac{5}{8}-\frac{5}{48}} \nonumber 
 \left|g'_A(z_0) \right |^{\frac{5}{48}}
\end{eqnarray}
The correct definition of the square root is:
\begin{eqnarray}
 \sqrt{x+i y}&=&\frac{1}{\sqrt{2}}(  \sign(y) \sqrt{r+x}  + i \sqrt{r-x}) \\
 r &=& \sqrt{x^2+y^2}
\end{eqnarray}
Hence the true map $g_A(z)$ reads for $z=x+i y$, $y \geq 0$:
\begin{eqnarray}\nn
 g_A(z) &=& \frac{1}{\sqrt{2}}(  \sign(x-a) \sqrt{\rho+t}  + i \sqrt{\rho-t})  \\
&&  + {\sign}(a) \sqrt{a^2+h^2} \\
 \rho &=& \sqrt{((x-a)^2-y^2+h^2)^2+4 (x-a)^2 y^2}\qquad  \\
 t &=& (x-a)^2-y^2+h^2
\end{eqnarray}
One finds:
\begin{eqnarray}
 |g'_A(z)|^2 &=& \frac{(x-a)^2+y^2}{\sqrt{h^4 + 2 h^2 ((x-a)^2-y^2)+((x-a)^2+y^2)^2}}\nn\\
 & =& \frac{(x-a)^2+y^2}{\rho} 
\end{eqnarray}
%Finally: Pierre
%\begin{eqnarray}
%&& p=|\frac{a^2}{a^2 + h^2}|^{\frac{5}{16}}  ((x-a)^2+y^2)^{5/96} \rho^{-5/96} 
%(\frac{\rho-t }{2 y^2})^{25/96} \\
%&& \times (x^2+y^2)^{5/8} (\rho + a^2 + h^2 + \sqrt{2} \sign(x-a) \text{sign}(a)  \sqrt{\rho+t})^{-5/8}
%\end{eqnarray}
%Kay:
\begin{align}
{\cal P} = \left|\frac{a^{5/8} z_0^{5/4}
   \Im(g_A(z_0))^{25/48}
   g_A'(z_0)^{5/48}}{\left(a^2+h^2\right)^{5/16}
   y_0^{25/48} g_A(z_0)^{5/4}}\right|
\end{align}
This simplifies considerably for $x=a$
\begin{align}
p= \left(\frac{a^2}{a^2 + h^2} \right)^{{5}/{16}} \left(\frac{y^2-h^2}{y^2}\right)^{5/24} \ ,
\end{align}
which is the result given in the main text.
%\newpage{\hspace{4cm} }
%\newpage

\section{Appendix B: Force on a disk}

Let us recall \cite{bauer} that the
probability that a SAW from $1$ to $e^{i \phi}$ on the unit disk avoids a disk centered at $0$ of radius $\rho\ll 1$ is
\begin{equation}
{\cal P}(\rho,\phi) \approx 1 - c \rho^{2/3} \sin^2(\phi/2) \ ,
\end{equation} 
where $c$ is still elusive. 
%Note \footnote{{\red what is r???}
%For a Brownian the hitting probability $G$ is solution of $\nabla^2 G=0$ with $G=1$ on the small disk and $G=0$ on the large one. The solution is $G=\ln r/\ln \rho$, hence for a typical starting point $G \sim 1/\ln \rho$ in agreement with the $d-d_f$ exponent.} that $2/3=d-d_f$. 
Consider the map $z=z_0 \frac{1-w}{1 - \frac{z_0}{\bar z_0} w}$, equivalent to ${w=\frac{\bar z_0}{z_0} \frac{z_0-z}{\bar z_0-z}}$ from the upper half plane in $z$, to the unit disk in $w$, with $w(0)=1$, $w(z_0)=0$ and $w(\infty)=\bar z_0/z_0=e^{i \phi}$.  We note $z_0=x_0+ i y_0=r_0 \rme^{i \theta_0}$ with $\theta_0=-\phi/2$. One finds that the circle of radius $\rho$ centered at $w=0$ is mapped to a circle in the upper half plane of center $z_c=x_c+ i y_c$ with $x_c=x_0$ and $y_c=y_0 \frac{1+\rho^2}{1-\rho^2}$ and radius $R=2 \rho y_0/(1-\rho^2)$. Hence $y_0=\sqrt{y_c^2-R^2}$ and the no-hit probability of the circle in the upper half plane is
%\begin{align}
%{\cal P}^{\text{half-plane}}=  \textstyle{\cal P}\left(\rho=\frac{y_c- \sqrt{y_c^2-R^2}}{R},\,\sin^2(\frac{\phi}{2})=\frac{y_c^2-R^2}{x_c^2+y_c^2-R^2}\right) 
%\end{align}
\begin{align}\label{40}
{\cal P}^{\text{half-plane}}\approx 1- c\left[\frac{y_c- \sqrt{y_c^2{-}R^2}}{R}\right]^{\frac23} \frac{y_c^2-R^2}{x_c^2+y_c^2-R^2}
\end{align}
For small $R$, $\rho \approx \frac{R}{2 y_c}$, $z_c = r e^{i \theta} \approx z_0$ hence $\theta=\theta_0=-\phi/2$ one finds the formula given in the text.

Another interesting limit studied in \cite{bauer} is $y_c-R \ll R$. There the no-hit probability in the unit disk for $0 \geq \phi \geq \pi$ is
\begin{equation}
{\cal P}(\rho,\phi) \approx \exp\left(- \frac{5 \pi}{8} \frac{\phi}{1-\rho}\right) \ .
\end{equation}
 In that limit, the leading free energy is
\begin{equation}
{\cal F} \approx k T \frac{5 \pi}{8} \phi \sqrt{\frac{R}{2(y_c-R)}} 
\end{equation}
with $\phi=\phi(x_c^2/(2 R (y_c-R))$,  $\phi(v)=2 \text{arcsin}{ \big(\sqrt{1/(1+v)}\big)}$, hence $\phi(0)=\pi$ and $\phi \sim 2 \sqrt{2 R (y_c-R)}/x_c$ for large $x_c$. Along the symmetry direction $x_c=0$  the force is
\begin{equation}\label{43}
f_y \approx k T \frac{5 \pi^2}{16} \frac{\sqrt{R}}{\sqrt{2} (y_c-R)^{3/2}} \ .
\end{equation}
%which is also the expression of the radial force in the other limit.
Note that the exponents of the $y_c$ dependence  matches $-3/2$ for  $y_c-R\ll R$, see Eq.\ (\ref{43}) and $-5/3$ at  $R\ll y_c$, see Eqs.\ (\ref{40}) and (\ref{14}).

%%%%%%%%%%%%%%%%%%%%%%%%%%%%%%%%%%%%%%%%%%%%%%%%%%%%%%%%%
%%%%%%%%%%%%%%%%%%%%%%%%%%%%%%%%%%%%%%%%%%%%%%%%%%%%%%%%%
%%%%%%%%%%%%%%%%%%%%%%%%%%%%%%%%%%%%%%%%%%%%%%%%%%%%%%%%%
%%%%%%%%%%%%%%%%%%%%%%%%%%%%%%%%%%%%%%%%%%%%%%%%%%%%%%%%%
% Uncomment if needed 
%%%%%%%%%%%%%%%%%%%%%%%%%%%%%%%%%%%%%%%%%%%%%%%%%%%%%%%%%
%%%%%%%%%%%%%%%%%%%%%%%%%%%%%%%%%%%%%%%%%%%%%%%%%%%%%%%%%
%%%%%%%%%%%%%%%%%%%%%%%%%%%%%%%%%%%%%%%%%%%%%%%%%%%%%%%%%
\end{document}